\documentclass[12pt]{article}
\usepackage{amsmath}
\usepackage{cite}
\usepackage[unicode, pdftex]{hyperref}

\begin{document}

\title{\textbf{Interpretation of crystallographic groups under Riemann's elliptic geometry}}

\maketitle
\begin{center}
A.P. Klishin, S.V. Rudnev\\ aklishin@yandex.ru, rudnew.stanis@yandex.ru \\
\textit{Department of Silicate and Nanomaterials Technology, Institute of High Technology Physics, TPU, 30 Lenin Avenue, Tomsk, Russia}
\end{center}

\textbf{Abstract}.This paper is devoted to the problem of choosing the most suitable model of a geometrical system for describing the real crystallographic space. It has been shown that all 230 crystallographic groups used to describe the crystalline structures in a Euclidean space can be presented by elliptic motions in the closed space $V^3$. 
Based on these results, it is stated that a special geometric system---the crystallographic space of interpretation $R_E$, determined by a form of an interpretant (the surface of a torus $T^2$ can serve as this interpretant)---can serve as a geometrical model of the real crystallographic space. 

The compact model of the closed structure of a crystal has been proposed and ways of its treatment for visualizing the constructions of elements of symmetry of a crystalline lattice in the Euclidean space $E^3$ have been determined. As a modeling space for describing the internal structure of a crystal, the closed space $V^3$ with the elliptic metrics and constant positive Gaussian curvature ($K=1$) has been offered. The properties of the internal space of a real crystal are naturally deduced from the properties of the modeling space.


\begin{center} \textbf{Introduction} \end{center}

Now the concept of non-Euclidean spaces is widely used to describe the general evolutionary principles of different physical systems ~\cite{bib:b1,bib:b2,bib:b3,bib:b4,bib:b5,bib:b6}. In this regard a genuine research interest in understanding the possible realizations of crystallographic groups in non-Euclidean spaces becomes perceptible. The realizations of crystallographic groups of symmetry were considered in a pseudo-Euclidean plane $R_1^n$ ~\cite{bib:b7,bib:b8,bib:b9}, Lobachevsky space $H^n$  \cite{bib:b10,bib:b14}, a Minkowski space $R_s^n$  ~\cite{bib:b12,bib:b13,bib:b14,bib:b16,bib:b17}, and the $n$-dimensional ($n>3$) Euclidean space $E^n$  ~\cite{bib:b18,bib:b19}. A major shortcoming of such considerations of crystal lattices in these spaces is that they use an infinitely extended space to simulate lattices. The crystalline structure of an ideal crystal can be built by multiplication of a finite number of atoms by all the transformations of any crystallographic group in a Euclidean a space. But such a construction of an ideal crystal is not connected with the natural reasons of crystallization, its finite character, and the form restricted in the space. Besides, the question on the relationship of an internal structure and an external faceting of a real crystal remains unresolved  ~\cite{bib:b20}. The ambiguity of geometrical interpretation of the space of the real crystal of minerals has led to the necessity to develop the models with the use of a non-Euclidean way of the description of crystalline structures.

At the heart of the offered theoretical approach lie the investigations of various non-Euclidean ways of the description of elements of a crystalline lattice  ~\cite{bib:b21,bib:b22,bib:b23,bib:b24,bib:b25,bib:b26,bib:b27,bib:b28}. When modeling the crystalline structures, the space $V^3$ of Riemann's geometry with constant Gaussian curvature $K=1$, which in infinitesimal domains coincides with the Euclidean space, is chosen as the modeling space. The crystalline structure is formed according to a certain crystallographic Fedorov group ($\Phi$) acting in the Riemann space ($F$). In this case we deal with a finite space where distances between any two points do not surpass some certain value ~\cite{bib:b29,bib:b30,bib:b31,bib:b32,bib:b33}. For crystallographic structures in an elliptic space, as well as in the case with a Euclidean space, the conditions of global discreteness and homogeneity are satisfied. 

It is known that not all the non-Euclidean spaces suppose embedding in a Euclidean space of an acceptable dimension ~\cite{bib:b34}. In this work, the modeling representations of crystalline structures, which suppose an acceptable interpretation of images of the elliptic space in the form of representations in a Euclidean space, are considered. Thus, the problem of construction of the interpretation of a closed restricted manifold which describes a lattice crystalline structure is reduced to the construction of the special representations, the corresponding crystallographic groups acting in a Riemann's elliptic space. Under the conditions of Riemann's geometry, it will be natural to preserve the term “crystallographic group” for the discrete groups of motions, which we will denote as $F$-groups. 

The basic difference from existing approaches to modeling consists in the statement, according to which the organization of a lattice structure occurs according to a certain $F$-group ($I_E^R (F)=\Phi$, where $I_E^R$ is a functional establishing the correspondence between the crystallographic (Fedorov) group $\Phi$ in the Euclidean space $E^3$ and the $F$-group acting in the Riemann space $V^3$). 

Modeling a crystalline structure in a Euclidean space has the principal difference from the approach proposed in this paper and consists in the following: in a Euclidean space they form the ideal model of a crystal, which possesses a translational symmetry, has an infinitely extended lattice, and does not possess a center of symmetry, zonality, and sectoriality. A description of different types of locations and shapes of faces limiting the crystal surfaces in the traditional models is solved by introducing the different artificial boundary conditions which are very weakly correlated with the conditions of the evolution of crystals. The ideal Euclidean model of a crystal does not possess those structural features that characterize real crystals, except for a fragmentary coincidence to the restricted regions of a spatial lattice.

In this paper, the ideal models of crystals and crystalline microstructures are restricted in sizes, have a certain form and symmetry, possess the properties of zonality and sectoriality, and also possess the centers of symmetry. The above properties constitute a basic set of structural characteristics that are present in real crystals of minerals.
 
\section{ Interpretations of crystallographic groups under the conditions of Riemann's elliptic geometry}

It is supposed that the organization and evolution of the lattice structure of a crystal occurs in the Riemann's elliptic space $V^3$, and the crystalline structure is formed according to a certain crystallographic group acting in the Riemann space. An atom in a lattice system model is considered as a point node, and the point systems are studied at the initial stage without chemical features of that atom. Owing to the traditional representations, the visualization of the lattice structures obtained is carried out in the Euclidean space.

For further consideration it is necessary to introduce a number of definitions.

As a lattice system we will understand a geometrical model of the structure of a crystal, determined generally by a set of the crystallographic groups ($\Phi_1,\Phi_2\ldots$) in the Euclidean space or in the Riemann's geometry space \\ ($F_1,F_2, \ldots$).

The group of motions of a torus  $H_S$ represents a group of transformations of the surface of a torus relative to the mutually perpendicular axes of symmetry and consists of the two subgroups: translational $H_T^t$ and orthogonal $H_T^0$, i.e. $H_T=H_T^t \times H_T^0$. With the help of this group, the construction of $R_E$~-~interpretation of a lattice structure on the basis of images of the Euclidean space is carried out.

The space of  $R_E$-interpretation is a set of points of the Euclidean space (a toric manifold), where the model of a lattice system of a crystal ($\Phi_1,\Phi_2\ldots$) is described by means of the groups $\{H_T (\Phi_1 ),H_T (\Phi_2 ),\ldots\}$ of motions of a torus $H_T$. The space of $R_E$-interpretation has the flat Euclidean metric and is closed and bounded.

The group of motions of a Clifford surface $H_S$ is a group of transformations of the surface of $S_K$, consisting of the two subgroups: translational $H_S^\tau$ and orthogonal $H_S^0$, i.e. $H_S=H_S^\tau\times H_S^0$. With the help of this group, the description of the structure of a lattice system ($F_1,F_2,\ldots$) in the Riemann's geometry space $V^3$ ($K=1$) is carried out. According to S.V. Rudnev's basic postulate ~\cite{bib:b24}, the crystallographic space of a crystal satisfies to the axioms and theoretical proposition of Riemann's elliptic geometry ($T_R$). 

The $R_E$-interpretation consists in comparison to all the initial notions of the theory $T_R$ of certain mathematical objects and relations between them, some system in $R^3$, where $T_R$ is the axiomatic theory of construction of Riemann's geometry, proposed by Bogomolov ~\cite{bib:b35}. Further we will speak about the interpretation $T_R$ in the geometry $R^3$, and the system $R_E$ we will call as the interpretation of the theory $T_R$. In application to the crystallographic geometry, only a special case of the interpretation of lattice systems is considered. In this case, the interpretation of the lattice systems which are taking place in the space of Riemann's geometry is carried out by means of images of Euclidean three-dimensional geometry. As a base element in Riemann's geometry, the Clifford surface has been chosen, whose geometrical image in the Euclidean space is the torus.

The general scheme of the interpretation of Riemann's elliptic geometry in the Euclidean space ($I_E^R$) is simple enough.

A certain geometrical image $A$ of the Riemann's geometry space, $A\in V^3$, possessing the necessary properties, is chosen as an object of the interpretation in the Euclidean space, by means of a some geometrical image $B$ of the Euclidean space, $I_E^R(A)=B$, where $B\in R^3$. The chosen geometrical image $B$ of the Euclidean space is allocated with the properties of the geometrical image $A$ and thus becomes a carrier of the properties of the geometrical image of the Riemann space in the Euclidean space. 

Operations with the geometrical image $B$ are carried out by rules of the Riemann space, taking into account the certain conditions stated below. As a result, in the Euclidean space there appears a certain region $R_E$ which is called the space of interpretation. We have 
\begin{equation}\label{eq1}
I_E^R (M)=R_E
\end{equation}   
where $M\in V^3$ is area of the Riemann space which is subjected to the interpretation.

As a basic geometrical element of the Riemann's elliptic space for constructing the space of interpretation $R_E$ the Clifford surface $S_K$---direct circular cylinder of the elliptic space---is chosen. Isometric embeddings of a flat torus in the three-dimensional elliptic space were first found by W. Clifford; this was the starting point and basis for constructing the interpretation of the elliptic space using Euclidean forms ~\cite{bib:b44}.The most important thing in justifying the choice of the basic geometric element of the interpretation is the fact that Euclidean geometry $E^2$ takes place on $S_K$.

The surface $S_K$ is isomeric to a Euclidean rectangle or rhombus with the identified opposite sides, what leads us to an ordinary Euclidean torus. Then
\begin{equation}\label{eq2} 
I_E^R (S_K )=T^2
\end{equation}  
where $T^2$ is a $2D$ torus belonging to the Euclidean space $R^3$.

However with the principal difference that in the space of interpretation the torus loses its geometrical independence and becomes the carrier of properties of $S_K$, i.e. in the space of interpretation the torus is just the surface on which the Euclidean geometry is fulfilled and well-known groups of motions of a Euclidean torus are more not applicable to it. To operate with a torus in the space of interpretation we should follow the other rules. In the space $R_E$ the torus motions are rotations. At the same time, using that $R_E$ "is located" in the Euclidean space, it is possible to carry out the sections of $R_E$ by a Euclidean plane and in such a way to study the properties of structures in the space of interpretation. It is extremely important that the Clifford surface is the carrier of Euclidean geometry, then in $R_E$ on a torus (or on systems of tori) all the physical laws prove themselves in the same form and with the same content as on an Euclidean plane and in space, though, in a bit different form. The difference is that instead of a segment of a Euclidean straight line the arc lying on a torus is used. In Coulomb's law in $R_E$  , for example, $l$---an arc on a torus---is used in place of $r$ .

The relative simplicity of manipulations with the crystallographic groups in the elliptic Riemannian space begets a complexity of constructing the geometrically clear picture of their images in three-dimensional Euclidean space; here we mean the lattice systems representing the structure of a real crystal. In order to solve this problem, the interpretation method is used.

Images of translations on a Clifford surface in $R_E$ are expressed in rotations of a torus about own axes and determined by the group $H_T~=~\{H_{T}^{l_1},H_{T}^{l_2}\}$, where $H_{T}^{l_1}$ is the subgroup of rotations about axis $l_1$, and $H_T^{l_2}$ is the subgroup of rotations about axis $l_2$, that is, in sliding on itself. We have  
\begin{equation}\label{eq3}
I_E^R(H_S )=H_T
\end{equation}  
where $H_S$ is the translational group generated from subgroups of the paratactic shifts of $S_K$.

Translations of the Clifford surface itself in Riemann space are expressed in $R_E$ in rotations of a torus about own generatrices. Then
\begin{equation}\label{eq4}
I_E^R(H_S(S_K))=H_T(T^2 )
\end{equation}  
Different crystallographic axes of symmetry (from $l_2$ to $l_6$) are considered on different Clifford surfaces, what in $R_E$ is expressed in application of tori with a different relation of internal and external radiuses. When unfolding the tori it is shown in the form of rectangles with different lengths of the sides and different angles between their diagonals (for axis $l_3$, for example, the angle between diagonals is taken to be equal to $60$ or $30$ degrees, for axis $l_4$ this angle is equal to $45$ degrees, with corresponding lengths of the sides of the rectangle unfolded). Since
\begin{equation}\label{eq5}
I_E^R(H_S^{m_i}(S_K))=H_T^{l_i}(T^2 (r_i,R_i))
\end{equation}
where $r_i$,$R_i$ are the internal and external radii of a torus; $H_T^{l_i}$ are the subgroups of rotation of $T^2$; $H_S^{m_i}$ are the subgroups of rotation of $S_K$; $i$ is the order of symmetry. For we have    
\begin{equation}\label{eq6}
I_E^R{(O_S)}=O_T
\end{equation}
where $O_T=\{H_T{T_0^2}(r_k,R_k))\}$ is the subgroup of rotations of a torus, and the relation $r_k n=R_k m$ will be true for the torus radiuses.

The three-dimensional space of interpretation $R_E$ consists of a system of tori of Euclidean space, which are connected with each other by relations of the same order as the corresponding Clifford surfaces of elliptic space. The images of surfaces with the Euclidean metric in a sector of the three-dimensional space of interpretation are tori. The rest of the space of interpretation describes representations of the Riemann's elliptic space. Thus, it is possible to describe the space of interpretation through its forming groups:
\begin{equation}\label{eq7}
R_E=\{H_T,O_T\}
\end{equation}
where  $H_T$ is the translational subgroup of a torus, and $O_T$ is the subgroup of rotations of a torus. It is easy to show that the space of interpretation is restricted, closed, and continuous. Different classes of symmetry of a crystalline structure when modeling will be determined by the value of the relation of internal and external radiuses of a torus. 

It is convenient to study the geometrical and structural features of objects in $R_E$ by means of sections of the space of interpretation by Euclidean planes. However, the most important property of the space of interpretation is that on a torus in $R_E$ we have the right to consider physical laws in their usual Euclidean interpretation, with the Euclidean segments ($r$) replaced with the arcs of the torus ($l$). 

\subsection{General formulation of the problem. Discrete groups of motions of spaces of positive curvature}

The theory of discrete groups of motions of spaces of constant curvature attracts attention of many researchers ~\cite{bib:b37,bib:b38,bib:b39,bib:b40} and is a part of the general theory of discrete Lie groups. In crystallography, the groups of symmetries of crystalline structures were traditionally considered, being the discrete subgroups of groups of motions of the three-dimensional Euclidean space $E^3=R^3\times O(3)$.

A Lie group of the three-dimensional Euclidean space $E^3$ is disconnected, whose connected component $E_0^3$ consists of own motions $SO(3)$ (i.e. preserving the orientation). There is the following Levi decomposition for the group of own motions: $E_0^3=R^3\times SO(3)$. The normal subgroup $R^3$ consists of parallel translations. The lattices $\Gamma$ in $E^3$ are called the crystallographic groups. The population $L=\Gamma \cap R^3$ of all translations from $\Gamma$ has a finite index in $\Gamma$, and the factor group (the group of linear parts) $\Phi=L/\Gamma\subset O(3)$ is finite. A set of motions in the Euclidean group of three-dimensional space, by which all the crystallographic groups are formed, can be presented as follows: 
\begin{equation}\label{eq8}
\alpha (x): x\mapsto gx+t(g)
\end{equation}
where $x\in E^3$, $g\in\Phi$. 	

The vector $t(g)$  is defined accurate within an addition from $L$. As in the two-dimensional case, the mapping $\alpha   \mapsto g$ determines a surjective homomorphism $Isom(E^3 )\mapsto O(3)$, whose kernel, as it is known, coincides with the group of shifts of the space $E^3$. 
  
The geometrical description of the space $E^3$ differs from the corresponding description in two dimensions. For example, any orientation-preserving isometry of the space $E^3$ is a screw motion representing a composition of shift and rotation around a straight line invariant relative to this shift. Thus both shifts and rotations are special cases of isometries of this type.

It is known that if $G$ is a discrete group of isometries of the plane $E^2$ (in our case, a plane crystallographic group), then its subgroup of shifts has in it a finite index. And the same is true for n-dimensional Euclidean space; in this case there is a free Abelian subgroup of a finite index, whose rank does not surpass $n$ (Bieberbach's theorem) ~\cite{bib:b46}. Using this result, it is easy to show that either the subgroup of shifts of the group $G$ has in it a finite index or $G$ is a finite expansion of the group $Z$. 

With the approach proposed by us, for describing the structural features of a crystalline structure, the closed model of a crystal is used, and the discrete groups acting in the elliptic space $V^3$ must act freely and not have a torsion. Used in this work, a modeling manifold is closed, and the groups acting on it as isometries of $V^3$ must act freely and not have a torsion, as in the Euclidean case. 

Suppose that H is the orientation preserving subgroup $O(3)$---that is, a subgroup in $SO(3)$. Use the last as a translational component for describing the actions of crystallographic groups in elliptic space. From the complete classification of subgroups of the group $SO(3)$ it follows according to ~\cite{bib:b41} that H can be a cyclic group, or a dihedral group, or the group of orientation preserving symmetries of one of five regular polyhedra. 
Discrete regular systems are possible not only in Euclidean space, but also in some spaces possessing a positive Gaussian curvature. One of confirmations of the possibility of atomic constructions with the non-Euclidean metric is fullerenes which can be considered as ideal crystals of spherical space ~\cite{bib:b11}. 
  
The spherical (elliptic) space of constant positive Gaussian curvature takes a special place in geometry along with Euclidean and Lobachevskii spaces. This space is the most accessible to our geometrical intuition, what allows us to use it for constructing the closed, finite model of the structure of a crystal.

\textbf {General formulation of the problem of presentation of crystallographic groups in the elliptic space} 

\textit{The presentation of crystallographic groups in the elliptic space can be represented as a particular case of the Nielsen realization problem}  ~\cite{bib:b38}\textit{, follows: find and describe the realization of $230$ Fedorov crystallographic groups $(\Phi_1,\Phi_2 \ldots)$ using the homeomorphisms $f_F$  of the elliptic space in such a way that the last form a group of transformations of a surface $S$ (in our case, the role of $S$ plays the three-dimensional spherical space $S^3$).} 

The given formulation leads to the equivalent problems of constructing the realizations of the crystallographic groups  $(\Phi_1,\Phi_2 \ldots)$ by subgroups of the group of:

\begin{enumerate}
\item \textit{Spherical diffeomorphisms of a manifold of $S^3$;}
\item \textit{Motions of the elliptic space $V^3$;}
\item \textit{Motions (collineations) of the space $RP^3$.}
\end{enumerate}

From work ~\cite{bib:b41} it is known that for the case of spherical (elliptic) spaces the problem in the general formulation has a positive solution. 

\subsection{Fundamental domains of action of groups \\ of constant curvature }

It is known from ~\cite{bib:b39} that any discrete group of motions of spaces of positive curvature possesses a convex fundamental domain. As a plane finite group of motions is discrete, any finite group of motions of the space $E^3$ has a fixed point. Therefore, such a group can be considered as a discrete group of motions of sphere with the center in this point. Further, suppose that the fundamental domain of the group $\Gamma$ is a closed region $D\in V^3$ such that a subset $\{\gamma D,\gamma\in\Gamma\}$ forms a partition of the modeling space $V^3$.
	
In the majority of considered cases, the fundamental domain manages to be chosen in the form of a convex polyhedron with a finite number of facets. When moving from the group $\Gamma$, the fundamental domain passes into regions which are equivalent to it and all the set of images of the fundamental domain forms a space partition, and the group $\Gamma$ acts transitively on this partition.

In geometrical crystallography of Euclidean space, the discrete groups of some special kind are usually considered, which are called Fedorov ($\Phi$) or spatial crystallographic groups. As a crystallographic group it is accepted to call the symmetry transformations’ group acting transitively on some regular point system that possesses the following properties ~\cite{bib:b43}.

\begin{enumerate}
\item There is at least one point $A$ of space, isolated in a class $A_\Phi =\{f(A)\mid f\in\Phi\}$ of points equivalent to it, i.e. there is a sufficiently small radius $r$ called the radius of discreteness of point $A$ such that a sphere $\omega =\omega(r,f(A))$ of this radius having the center in a point $f(A)$ of an orbit of the point $A$ contains a unique point of this orbit---the sphere’s center (local discreteness).
\item There is at least one such a point in space that the class of points equivalent to it is located in space uniformly, i.e. there is a sufficiently large radius $R$ called the radius of homogeneity of the point $B$ such that in any sphere $\omega$ of this radius there will be a point equivalent to $B$ (local homogeneity).
\end{enumerate}
	
The conditions 1 and 2 imposed on a crystallographic group are quite natural and open the possibilities for constructing the actions of the groups in the modeling elliptic space due to its topological formulation. From the local homogeneity the global homogeneity follows, namely the existence of a sufficiently large radius $R^{'}$  such that in any sphere of this radius there will be a point equivalent on the group $\Phi$ to any preassigned point of space (thus  $R\leq R^{'}$) ~\cite{bib:b44}. From the global homogeneity and local discreteness the global discreteness of an orbit of any point follows. From the given conditions 1 and 2 of crystallographic groups, and the theorems of Fedorov \cite{bib:b20,bib:b38,bib:b39,bib:b40} and Schoenflies-Bieberbach ~\cite{bib:b46} it follows that it is possible to establish an interpretative correspondence between the elliptic motions of space $V^3$ \\ ($F$-groups) and the crystallographic groups of Euclidean space (Section \ref{s2}). In Riemann's geometry $V^3$, it is logical to preserve the term "Fedorov group" for the motions’ discrete groups satisfying to the conditions for a regular system of points.

The fundamental domains of discrete groups of transformations can be constructed upon very broad assumptions concerning the modeling space. If $F$ is a discrete group of isometries of space of positive curvature, then as its fundamental domain it is possible to choose a Dirichlet region: 
\begin{equation}\label{eq9}
D_{x_0}(F)=\{x\in R^3 \mid \rho (x,x_0)\leq \rho (\gamma x,x_0 ),\forall\gamma\in F\}
\end{equation}
where $x_0$ is any point whose stabilizer is trivial, $\rho$---flat Euclidean metric. 

A lack of the Dirichlet fundamental domain is a special role which an infinitely remote point plays. This lack eliminates a construction of the fundamental domain of some special kind---a fundamental polyhedron of the discontinuous group $F$ with the center in any point of a set of discontinuity ~\cite{bib:b40}. 	

\subsection{Groups of motions of elliptic geometry. \\ A translational group of Clifford translations  \\ (paratactical shifts)}

Any orthogonal transformation $g$ of Euclidean space $R^{n+1}$ in a suitable orthonormalized basis is written in the form of a cellular diagonal matrix $g=(1,\ldots 1,-1\ldots -1,R(\phi_1)\ldots R(\phi_s ))$, where 
\begin{equation}\label{eq10}
R(\phi)=
\begin{bmatrix}
 \cos(\phi)  & \sin(\phi) \\ -\sin(\phi) & \cos(\phi) 
\end{bmatrix}
\end{equation}
is the matrix of rotation through an angle $\phi$, $0<\phi<\pi$.

Let $V(\phi )$ denote the direct sum of subspaces, answering to the diagonal blocks $R(\phi_i)$,  with $\phi_i=\phi$. The transformation $g$ induces in $V(\phi )$ a complex structure $I$ which acts in every $g$-invariant two-dimensional subspace $U~\in~V(\phi)$  as the rotation through the angle $\phi$, in the same direction as $g\mid_U$. The restriction of the transformation $g$ on the subspace $V(\phi )$ looks like 
\begin{equation}\label{eq11}
g\mid_{V(\phi)}=E \cos(\phi)+I \sin(\phi)
\end{equation}
and induces in a plane of the sphere $V(\phi)\cap S^n$ the Clifford translation with the value of displacement $\phi$. Then we use the following definition of the axis of motion as a nonempty critical set $M_g^\lambda$ of the function of displacement $\rho_g$. It is necessary to note that for the case $S^3$ any motion possesses, at least, one axis.
If $g$ is the motion of a sphere and $M_1,\ldots M_S$ are its axes, then the motion $g$ induces a Clifford translation $g_i=g\mid_{M_i}$ on each axis and expands into the product of commuting slidings along the axes $M_i$,  induced by these translations. Accurate within the conjugation in the group, the motion is uniquely determined by critical values of the function of displacement and dimensions of the axes.

Thus any one-parametrical group of motions of a sphere decomposed into the product of commuting one-parametrical groups which are canonical continuations of the one-parametrical groups $E\cos(t)+Isin(t)$  of Clifford translations of the pair of reciprocal polar axes ~\cite{bib:b39}.

The geometry on the sphere $S^3$ in the space $R^4$ possesses a strong resemblance to the geometry of the space $R^3$: there is a transitive group of one-to-one isometric motions, depending on 6 real parameters. However, if the group of motions of the space  $R^3$ is primitive, then this space cannot be broken into such sets in which at any motion each of these sets entirely passes in itself or in other of these sets. The group of rotations of the sphere is imprimitive: it can be broken into such sets, for example, the pairs of diametrically opposite points; at any rotation each of these pairs of points entirely passes in itself or each other. Further, if two straight lines of the space $R^3$ can cross only at a single point, then the large circles of spheres of the space $R^4$, playing a role similar to that of straight lines, intersect not in one point, but in two diametrically opposite points. For elimination of these mismatches, it is enough to identify diametrically opposite points, i.e. to consider a pair of diametrically opposite points of the sphere as a point of new space---the elliptic three-dimensional space $V^3$.

If the radius of $S^3$  is equal to $\rho$, then the space $V^3$ obtained by identifying diametrically opposite points is the Riemannian three-dimensional space with the curvature $1/\rho^2$. It is known that the spherical transformations of the space $V^3$ preserving distances between its points are rotations. These rotations form a group of motions of the space $V^3$. Owing to the definition of the space $V^3$, the group of its motions is primitive. 

From the fact that a sphere of the space $R^4$ is closed and compact, it follows that the space $V^3$ is closed and compact as well and is not divided by planes on two regions.

To the identical transformation of the space $S^3$ there correspond two motions---the identical transformation of the space $R^4$  and the mapping from the sphere’s center. The group of motions of the space $V^3$ is isomorphic to the factor group of rotations of the space $R^4$ on its subgroup of the antipodal mapping. At $n=2$ the group of motions of the space $V^2$ is isomorphic to a connected component of group of rotation of the space $R^3$. The motions of the space $V^3$, as well as the rotations in $R^4$, can be written as follows ~\cite{bib:b45}:
\begin{equation}\label{eq12}
x^{i^*}=U_j^i x^j
\end{equation}
where $U=(U_j^i)$ is an orthogonal operator satisfying to the orthogonality condition:

\begin{equation}\label{eq13}
\sum_i U_j^i U_k^i=\delta_{jk}
\end{equation}

In the modeling space considered, the determinants of the matrixes $(U_j^i)$ are equal to $\pm 1$, and the motions which answer them will be accordingly the motions of the first and second kind.

The transformations ~\cite{bib:b13} can be considered as collineations of the three-dimensional projective space $(P^3)$. Hence, the group of motions of the space $V^3$ is a subgroup of the group of collineations of the space $P^3$. 

As the group of motions ~\cite{bib:b12} of the space $V^3$ is isomorphic to a connected component of the group of rotations of the space $R^4$, it is possible to conclude that the group of motions of the space $V^3$ is a Lie group. Thus the classification of motions of the space $V^3$ is reduced to the classification of rotations of the space $R^4$. The simplest rotation of space is the rotation about 2nd plane; for the plane $x^0=x^1=0$  the rotation looks like:
\\
\begin{equation}\label{eq14}
\begin{cases}
x^{0^*}=x^0 cos(\phi)+x^1 sin(\phi)\\
x^{1^*}=-x^0  sin(\phi)+x^1 cos(\phi)\\
x^{2^*}=x^2\\
x^{3^*}=x^{3^*}
\end{cases}
\end{equation}

At this rotation all the points of the chosen plane are motionless, and in the 2nd plane $x^0 Ox^1$ the rotation through the angle $\phi$ occurs. At this rotation on $S^3$ with the center at the origin of co-ordinates, all the points lying on a 2-dimensional plane $x^0=x^1=0$ remain motionless, the large circle cut by $x^0 Ox^1$ passes in itself, and the other points move in small circles. Thus, all the points of the 2-dimensional plane $x^0=x^1=0$ remain motionless when rotating in the modeling elliptic space $V^3$, as well as a straight line defining some direction, and the other points move in circles. We will call such a motion, as well as in \cite{bib:b45}, the shift along a straight line. 

If the radius of curvature of the space $V^3$ is equal to $\rho$, to the rotation of the space $R^4$ through the angle $\phi$ there corresponds the shift of the space  $V^3$  on the distance $\delta=\phi\rho$, which can be presented as follows:
\begin{equation}\label{eq15}
\begin{cases}
x^{0^*}=x^0 \cos(\frac{\delta}{\rho})+x^1 \sin(\frac{\delta}{\rho})\\
x^{1^*}=-x^0  \sin(\frac{\delta}{\rho})+x^1 \cos(\frac{\delta}{\rho})\\
x^{2^*}=x^2\\
x^{3^*}=x^{3^*}
\end{cases}
\end{equation}

To the stationary angles $\phi_i$ of rotations of the space $R^4$ there correspond the stationary distances of motions of the space $V^3$, that are equal to $\delta_i=\phi_i \rho$; and when representing the motions of the space $V^3$ by the rotations of the space $R^4$, the stationary angles $\phi_i$ of rotations of the space $R^4$ should be replaced with the relation $\delta_i/\rho$, where $\delta_i$ are the stationary distances of motions of space $V^3$. Own vectors of rotations of the space $R^4$ determine the motionless points of motions of the space $V^3$, and the invariant 3-dimensional planes of rotations of $R^4$ determine the invariant 2-dimensional planes of motions of the space $V^3$. The invariant straight lines of motions of space $V^3$ are called as the rectilinear trajectories of the motions and serve as generatrixs of the translational part of actions of crystallographic groups in Riemann's elliptic space. 

On a plane $V^2$ with the radius of curvature ρ the matrix of rotation about a point looks like:
\begin{equation}\label{eq16}
\begin{vmatrix}
\cos(\frac{\delta}{\rho}) & \sin(\frac{\delta}{\rho}) & 0\\
-\sin(\frac{\delta}{\rho}) & \cos(\frac{\delta}{\rho}) & 0\\
0 & 0 & 1
\end{vmatrix}
\end{equation}
Thus on the plane $V^2$ the rotation about a point has one rectilinear trajectory which is a polar of the center of the rotation. The basic types of possible motions of the modeling space $V^3$ are presented in Table 1.
\\

\pagebreak

\begin{table}

\caption{Types of motions in the modeling space $V^3$}
\begin{tabular}{|p{6cm}|p{7cm}|}
\hline
Motion type & Matrix form \\
\hline
\vskip 1pt
1. Screw motions & 
\vskip 1pt
$\begin{vmatrix}
\cos(\frac{\delta_0}{\rho}) & \sin(\frac{\delta_0}{\rho}) & 0 & 0\\
-\sin(\frac{\delta_0}{\rho}) & \cos(\frac{\delta_0}{\rho}) & 0 & 0\\
0 & 0 & \cos(\frac{\delta_1}{\rho}) & \sin(\frac{\delta_1}{\rho})\\
0 & 0 & -\sin(\frac{\delta_1}{\rho}) & \cos(\frac{\delta_1}{\rho})
\end{vmatrix}
$\vskip 1pt
\\
\hline
\vskip 1pt
2. Rotation about a straight line & \
\vskip 1pt
$\begin{vmatrix}
\cos(\frac{\delta_0}{\rho}) & \sin(\frac{\delta_0}{\rho}) & ~~~~~0 & ~~~~~~~~~0~~~\\
-\sin(\frac{\delta_0}{\rho}) & \cos(\frac{\delta_0}{\rho}) & ~~~~~0 & ~~~~~~0\\
0 & 0 & ~~~~~1 & ~~~~~~1\\
0 & 0 & ~~~~~1 & ~~~~~~1
\end{vmatrix}$\vskip 1pt
\\

\hline
\vskip 1pt
3. Paratactical shift & 
\vskip 1pt
$\begin{vmatrix}
\cos(\frac{\delta}{\rho}) & ~\sin(\frac{\delta}{\rho}) & ~0 & 0\\
-\sin(\frac{\delta}{\rho}) & ~\cos(\frac{\delta}{\rho}) & ~0 & 0\\
0 & ~0 & ~\cos(\frac{\delta}{\rho}) & \sin(\frac{\delta}{\rho})\\
0 & ~0 & ~-\sin(\frac{\delta}{\rho}) & \cos(\frac{\delta}{\rho})
\end{vmatrix}
$\vskip 1pt
\\

\hline
\end{tabular}
\end{table}

1. The screw motion has two rectilinear trajectories which are polar straight lines. In this case (Type 1, Table 1), the axes of the directions $l_1$ and $l_2$ are shifted over the distances $\delta_1$ and $\delta_2$, respectively.

2. When turning around a straight line the motion has one rectilinear trajectory (Type 2, Table 1) in the direction of the axis $l_2$, and the angle of rotation $\delta/\rho$.

3. The paratactic shift has a 2-parameter set of straight trajectories (Type 3, Table 1). When rotating around the axis, the points of the space, which are not lying on the axis and on a straight-line trajectory, move in circles in planes perpendicular to the axis.

One of the central moments when constructing the interpretative correspondence is the establishment of correspondence between the translational group of parallel translations $E^3$ and the group of Clifford paratactic shifts (Type 3, Table 1). From properties of the paratactic shifts it follows that each point of the space $V^3$ moves on the same distance $\delta$ and each straight line, connecting an arbitrary point $x$ with the point $x^{'}$ into which the point $x$ passes at the paratactic shift, is a rectilinear trajectory of paratactic shift. Through each point of the space $V^3$ there passes one and only one rectilinear trajectory of paratactic shift.

The complete group $G=O(4)$ of motions of the sphere $S^3$ is compact. This determines a number of features of the groups of motions of the sphere in comparison with the groups of motions of other spaces of constant curvature. Note that there are many irreducible motions of the sphere: any irreducible orthogonal representation $\phi:H\rightarrow O(4)$ of the arbitrary compact group H determines a irreducible group $\phi(H)$ of motions of the sphere $S^3$. On the other hand, there exists, accurate within a conjugation in the group $G$, only a finite number of the coherent transitive groups H of motions of the sphere $S^3$ (Table 2).

\begin{table}

\caption{Connected groups of motions of sphere $S^3$}
\begin{tabular}{|l|l|l|l|l|l|l|}
\hline

Group ($H$) & $SO(3)$ & $SU(2)$ & $Sp(1)Sp(1)$ & $SO(2)Sp(1)$ & $Sp(1)$\\
\hline
Stabilizer ($H_x$) & $SO(2)$ & $SU(1)$ & $Sp(1)Sp(0)$ & $SO(2)Sp(0)$ & $Sp(0)$\\

\hline
\end{tabular}
\end{table}

It follows from the considered classification of motions (Table 1) that any own motion of the space $S^3$ is included in the one-parametrical group, as it is true for the Euclidean case $E^3$ as well. Any one-parametrical group of motions of the sphere is expanded into the product of commuting one-parametrical groups which are a canonical continuation of the one-parametrical groups $E\cos\lambda+I \sin\lambda$ of Clifford translations of some mutually polar axes. Any one-parametrical group of motions $R^4$ ($S^3$), not having the common motionless points, is either the product of a group of Clifford translations and a vector $\delta\in R$ or the product of some group and the one-parametrical group of rotations about a straight line with a directing vector.

An essential feature of elliptic geometry, unlike other non-Euclidean geometries, should be considered the existence of special motions of the space---the parallel shifts which force simultaneously all the straight lines of a given family of the parallels to slide on themself. The geometrical possibility of visual studying of crystalline structures with the help of elliptic motions consists in the representation of a group of Euclidean translations as a group of shifts. We shall consider Clifford translations as analogues of the parallel translations in the three-dimensional Euclidean space \cite{bib:b46}. 

The linear transformation $g\in O(4)$ is a Clifford translation if and only if either $g=\pm I$ or there exists an unimodular complex number $\lambda$ such that the transformation $g$ has a half of own numbers which are equal to $\lambda$ and a half equal to $\lambda$. In the case of elliptic space we will use exact orthogonal representations of the group $H$. If the group supposes a Clifford representation, then we shall say that the group H is a Clifford group. It is important to note that Clifford representations, as well as parallel translations, have no motionless points. 

On the structural structure $H$ is one of the following groups: cyclic, binary dihedral, binary tetrahedral, binary octahedral or, at last, binary icosahedral \cite{bib:b46}. 

\section{ The $I_E^R$-functional as a first approximation to the interpretative relation } \label{s2}

One of features of the spaces of constant curvature is their homogeneity so complete as well as that of a Euclidean space. This homogeneity is expressed in the existence of a group of motions with the same number of parameters as in a Euclidean space (for $n = 3$, $p = 6$; $n = 4$, $p = 10$). 

The following important feature is the free mobility of space, which allows us to conduct investigations by elementary-geometrical means \cite{bib:b47}. Other Riemannian spaces do not possess the free mobility; moreover, an arbitrary Riemann space, generally speaking, is absolutely not uniform and does not suppose any motions. When we consider the space $V^3$ as the sphere $S^3$ in $R^4$ with the identified diametrically opposite points, any motion of the space $V^3$ is determined by the sphere’s rotation. Thus the elliptic space is a sort of the  "folded double" spherical space. For construction of interpretation it will enough to us to show that all the crystallographic groups are realized in the spherical space on a hemisphere, whence at once their realizability will follow in the elliptic space.

Let $S^3$ be a sphere of the radius $R > 0$ with the center in the 4-dimensional space $R^4$ and $\Gamma$ be a group of 2nd order generated by the antipodal mapping $x\mapsto -x$. As it is the isometry of the sphere $S^3$ considered as a Riemann space with the metrics induced by the standard Euclidean metrics of the space $R^4$, then $RP^3=S^3/\Gamma$  possesses the unique Riemannian metrics, in relation to which the natural projection $S^3\mapsto RP^3$ is the locally isometric mapping. The projective space $RP^3$ supplied with this metrics is called the elliptic space.

In the case of a one-connected compact manifold, any locally-effective transitive action of a connected compact Lie group on $S^3$ is isomorphic to one of the following actions: to standard linear action $SO(4)$, $SU(2)$, $U(2)$, $Sp(2)$, $Sp(2)Sp(1)$. 

Interpretating the elliptic space $V^3$ as subspaces of the Euclidean space allows us to find a group of its isometries ($Isom$) $V^3$. Each orthogonal transformation of the space $R^4$ (a group element $O(4)$) translates in itself sphere $S^3$, so and $V^3$. Arising map $O(4)\mapsto isom S^3\mapsto isom V^3$ is a morphism that allows us to identify orthogonal transformations from $O(4$) with its image $isom V^3$.

Let's consider a generalization of the Bieberbach’s theorem  \cite{bib:b46}. Whether Let the $G$ connected group (elliptic motions), which semisimple part has no compact multipliers, is trivial acting on $G$, we will denote $N$ a nil-radical in $G$ Then if a $\Gamma$ lattice in $G$, a $\Gamma \bigcap N$ lattice in $N$.

As crystallographic groups form a Lie group which acts transitively and locally effectively on $S^3$, there is a nil-radical coinciding in this case with an Abelian radical. In particular, such groups are radical expansions of the standard action $SU(2)$ on $S^3$. This action is systatic, and it is possible to show that the group H of Clifford translations satisfies to this condition. Thus we have a locally effective action of the group $U(2)$ rad $H$ on $S^3$. For the group $G$ acting on $S^3$, in $G$ there is the simple normal subgroup $G_0$ transitively acting on $S^3$, and the following cases are possible: either $G=G_0$, or $G=SU(2)radG_0$, or $G=SO(3)radG_0$. From here, in the first case it is obtained for model pure translational group of Clifford translations, and in other cases set of the representations having both translational, and an orthogonal part. The Clifford parallelism on $R^4$ coincides with the Euclidean parallelism. We will denote through H a vector group of shifts (Clifford translations)
\begin{equation}\label{eq17}
t_\delta:y\mapsto y+\delta
\end{equation}
where $\delta,y\in V^3$, and through $O(4)$ a group of orthogonal transformations $R^4$ (the co-ordinate space). We will introduce a notation $E_3^s$---an image of Euclidean group in the space $V^3$, consisting of transformations having the following appearance:
\begin{equation}\label{eq18}
(g,t_\delta): x\mapsto gx+\delta(g)
\end{equation}
where $x\in R^4,g\in O(4)$, then $E_3^s$---the group of isometries containing actions of crystallographic groups in elliptic space, and $U(g)$---its subgroup, preserving motionless a point $O\in V^3$. Thus $E_3^s=U\times H$, consisting of everything $(g,t_\delta)\in U\times H$, and multiplication, as well as in a Euclidean case, is set by expression: 
\begin{equation}\label{eq19}
(g,t_\delta)(l,t_\gamma)=(gl,t_{g(\gamma)+\delta})
\end{equation}
The vector of paratactic shift $\delta(g)$ is determined to within addition. Thus, we establish following interpretative correspondence.

\begin{table}

\caption{The $I_E^R$-functional as a first approximation to the interpretative relation}
\begin{tabular}{| p{6cm}| p{6cm}|}

\hline
\begin{center}
 Euclidean space $R^3$
\end{center} &	\begin{center}
 Elliptic space $V^3$
\end{center}\\
\hline
$\Phi(g)$---an orthogonal subgroup, groups of rotations $R^3$ & $U(g)$---an orthogonal subgroup of rotations $R^4(S^3)$\\
\hline
$t(g)$---the parallel translation vector; the transformation has no motionless points & $\delta (g)$---a vector of the paratactic shift, which has not motionless real points\\

\hline
\end{tabular}
\end{table}

\subsection{Spherical manifold $S^3$}\label{s21}

The spherical manifold $S^3$ can be represented as follows:
\begin{enumerate}
\item[1)] a group of quaternions with the unit norm;
\item[2)] a set of the ordered pairs $(z_1,z_2)$ of the complex numbers, satisfying to the condition $|z_1|^2+|z_2 |^2=1$;
\item[3)] a unit sphere in $R^4$.
	
\end{enumerate}

Any three-dimensional manifold having geometrical structure on the sample $S^3$ should be oriented. Group of isometries of the manifold  $S^3$ allocated with the metrics, the induced Euclidean metrics $R^4$, the orthogonal group $O(4)$, therefore classification of actions on spheres serves is equivalent to classification of finite subgroups of the group $SO(4)$ freely acting on $S^3$. These groups was most completely were for the first time are classified by Hopf \cite{bib:b41}. For this classification isomorphism presence is important:
\begin{equation}\label{eq20}
(S^3\times S^3)/Z_2 \mapsto SO(4)
\end{equation}

The group $S^3$ acts on itself with the left and right multiplications, and it will be commuting isometries. Thus, the homomorphism is determined:
\begin{equation}\label{eq21}
\phi: (S^3\times S^3)\mapsto SO(4),									
\end{equation}
where $\phi(q_1,q_2)$ is the isometry of the sphere $S^3$  mapping $x$ in the $q_1 xq_2^{-1}$. 

The kernel of a homomorphism $\phi$ has the order 2, and a unique nontrivial element $(-1,-1)$. From this it follows that the image of mapping $\phi$ is a 6-dimensional group in $SO(4)$. As  $SO(4)$ is 6-dimensional and is connected, the mapping $\phi$ should be surjective, then $\phi: (S^3\times S^3)\mapsto SO(4)$. If $\Gamma$ is a subgroup of the group $SO(4)$ of the order 2 and freely acts on $S^3$ it consists of elements $I$ and $-I$ where through $I$ the individual matrix of the size $4\times 4$ is denoted. Then it is possible to say that at any finite subgroup G of group $SO(4)$, with free action on the $S^3$ order, either it is odd, or there is in accuracy one element of the second order which is equal $-I$.

Let's denote through $K$ a subgroup of the 2nd order ${I,-I}$ being the center of the group $SO(4)$. The group $K$  acts on  $S^3$ freely and the factor space on this action is the projective space $P^3$, therefore each element of group $SO(4)$ uniquely determines some isometry $P^3$. 

Thus, map $p:SO(4)\mapsto SO(3)\times SO(3)$ can be identified with natural mapping  $Isom(S^3)\mapsto Isom(P^3)$. Besides, we can identify $P^3$ with group $SO(3)$. The element $(u_1,u_2\in SO(3)\times SO(3)$ acts on $SO(3)$ mapping means $x\mapsto u_1 xu_2^{-1}$.

Each finite subgroup $SO(3)$  is one of the following: cyclic $Z_n$, dihedral $D_n$, tetrahedral $T$, octahedral $O$, or icosahedral $I$ group. If two finite subgroups of the group $SO(3)$ are isomorphic, they are conjugated in $SO(3)$ ~\cite{bib:b46}. 

If the finite subgroup $G$ of group $SO(4)$ freely acts on  $S^3$ the subgroup $H=p(G)$ of the group $SO(3)\times SO(3)$ freely acts on $P^3$. Therefore $H$ cannot contain any nontrivial element $(u_1,u_2)$ for which $u_1$, and $u_2$ are conjugated in $SO(3)$. We will consider more in detail, what finite subgroups $H$ of the group $SO(3)\times SO(3)$ can freely act on $P^3$. We will denote through $H_1$ and $H_2$ the projections $H$ to product the factors $SO(3)\times SO(3)$. Each of groups $H_1$ also $H_2$ is a finite subgroup in  $SO(3)$ and consequently represents a cyclic group, a group of the dihedron or a group of symmetries of some correct polyhedron preserving orientation.

\textbf{Proposition 1.} \textit {If $H$ is a finite subgroup of the group  $SO(3)\times SO(3)$ freely acting on $P^3$, and $H_1$ and $H_2$ are its projections to product factors, one of subgroups $H_1$, $H_2$ will be a cyclic one. If one of factors $H_1$ and  $H_2$ is not cyclic, one of following statements} ~\cite{bib:b46} \textit {is fairly:}

\begin{enumerate}
    \item[1)]\textit {$H=H_1\times H_2$, where orders of subgroups are mutually simple;}
    \item[2)]\textit {the noncyclical factor is isomorphic to group $T$, and $H$ is a diagonal subgroup of an index $3$ in $H=H_1\times H_2$; the cyclic factor has the order $3n$, where $n$ is odd;}
 	\item[3)]\textit {the noncyclical factor is isomorphic $D_n$ for odd $n$, and $H$ is a diagonal subgroup of an index $2$ in $H=H_1\times H_2$; the cyclic factor has the order $2m$, and $m$ and $n$ are mutually simple;}
	\item[4)]\textit {if  $H_1$ and $H_2$ are cyclic,then  $H$ will be a cyclic group.}
\end{enumerate}

The same approach is applicable also for a spherical case. Proceeding from a surjection of the mapping 
$\phi:(S^3 \times S^3) \mapsto SO(4)$ and an inclusion of a circle $S^1$ of complex numbers with the module 1 in $S^3$, with notations $\Gamma_1$ is a subgroup $\phi (S^1\times S^3)$ in $SO(4)$, $\Gamma_2$---$\phi(S^3 \times S^1)$ in $SO(4)$. Whence it follows that any finite subgroup $G$ of the group $SO(4)$ freely acting on $S^3$, is conjugated in $SO(4)$ some subgroup $\Gamma_1$ or $\Gamma_2$. Further we will present the exact description of all the possible subgroups $G$ of the group  $SO(4)$ which freely act on $S^3$. For any finite subgroup $\Gamma$ of the group $SO(3)$ we will denote through $\tilde{\Gamma}$ a prototype in $S^3$ rather the homomorphism $\phi: S^3 \mapsto SO(3)$, then   $\tilde{\Gamma}$ is the central expansion of group $\Gamma$ by means of $Z_2$. 

\textbf{Proposition 2.} \textit {Let $G$ be a finite subgroup $\Gamma_1$ freely acting on  $S^3$, then one of following statements for groups} ~\cite{bib:b46} \textit {is fulfilled:}
\begin{enumerate}
\item[1)]\textit {$G$ is cyclic;}
\item[2)]\textit {$H_2$ is isomorphic one of groups $T$, $O$, $I$ or $D_n$, and $H_1$ is the cyclic group, which order is mutually simple with the group order $H_2$; thus the group $G$ coincides with $\varphi (\tilde{H_1}\times\tilde{H_2})$ and is isomorphic $\tilde{H_1}\times\tilde{H_2}$;}
\item[3)]\textit {$H_2$ is isomorphic $T$, and the group $H_1$ is a cyclic one of  the order $3n$ with odd $n$, thus  $G$ is a diagonal subgroup of an index three in $\varphi (\tilde{H_1}\times\tilde{H_2})$;}
\item[4)]\textit {$H_2$ is isomorphic $D_n$, where $n$ it is odd, and the group $H_1$ is cyclic the order $2m$, and $m$ and $n$ are mutually simple; thus $G$ is a diagonal subgroup of an index two in $\varphi (\tilde{H_1}\times\tilde{H_2})$.}
\end{enumerate}

Hence, it is possible to conclude that $G$ is realized by a subgroup in $\varphi (\tilde{H_1}\times\tilde{H_2})$ and that $\tilde{H_2}$ acts on $S^3$ multiplication on the right, and $\tilde{H_1}$ multiplication on the left. We will consider, as before, the elliptic space as the factor space $S^3/\{\pm I\}$. In the case if the group freely transitively acts on $S^3$ and a subgroup $SO(4)$ freely acts on the factor space, thus, excluding antipodal images. It follows from proved before the statement for group action $H$---a finite subgroup of the group $SO(3)\times SO(3)$,  freely acting on $P^3$, where $H_1$ and $H_2$ are its projection to product factors.

\subsection{The representation of crystallographic groups \\ under the conditions of Riemann's geometry $V^3$}

One of the basic properties of the geometric model of ideal crystals is its invariance with respect to a finite number of orthogonal point transformations. We know 32 geometrical, crystallographic classes of point groups (Table 4) which are subsets of special orthogonal subgroups $SO(3)$, except for symmetry of the fifth order. The point groups have no translational components and they are easy for visualizing, using a projection to two-dimensional sphere from the center.

\begin{table}

\caption{Crystallographic classes}
\begin{tabular}{|l|l|l|l|l|l|l|l|}
\hline
$I \, C_i$        & $1 \, C_1$	  &         &          &        &          &        &	    \\
\hline
$2_m \, C_{2h}$   & $2 \, C_2$   & $m \, C_S$    &          &        &          &        &	    \\
\hline
$Mmm \, D_{2h}$    & $222 \, D_2$ & $mm2 \, C_{2v}$ &          &       &            &        &	    \\
\hline
$4mmm \, D_{4h}$   & $422 \, D_4$ & $4mm \, C_{4v}$ & $4i2m \, D_{2d}$ & $4m \, C_{4h}$ & $4 \, O_4$     &            & $4i \, S_4$ \\
\hline
$6mmm \, D_{6h}$   & $622 \, D_6$ & $6mm \, C_{6v}$ & $6i2m \, D_{3d}$ & $6m \, C_{6h}$ & $6m \, C_6$    &             & $6 \, C_{3h}$  \\
\hline
               &           &              &               & $6im \, D_{3d}$ & $32 \, D_3$   & $3m \, C_{3v}$ & $3i \, C_{31}$ \\
\hline
$m3m \, O_h$      & $432 \, O$    &             & $4i3m \, T_2$    & $m3 \, T_h$     & $23 \, T$     &              & $3 \, C_3$   \\

\hline
\end{tabular}
\end{table}

It has been found in \cite{bib:b48} that for investigation of crystals in Euclidean space, it is enough to consider twenty one algebraic transformations which are given in Table 5.

\begin{table}
\caption{Algebraic transformations forming crystalline classes}
\begin{tabular}{|l| p{12cm}|}

\hline

N & Algebraic transformations\\
\hline
$I$ &	$(x,y,z)\mapsto (x,y,z)$ identical transformation $\Delta=+1$\\
\hline
$X2$ &	$(x,y,z)\mapsto (-x,-y,-z)$\\
\hline
$X3$ & $(x,y,z)\mapsto (x,-\frac{1}{2} y+ \frac{\sqrt{3}}{2}z ,-\frac{\sqrt{3}}{2}y - \frac{1}{2}z)$, a cycling, rotation about axes $x,y,z$ on $180^{\circ}$, $120^{\circ}$, $90^{\circ}$, $60^{\circ}$, $\Delta=+1$\\
\hline
$X4$ & $(x,y,z)\mapsto (x,z,-y)$\\
\hline
$X6$ & $(x,y,z)\mapsto (x,-\frac{1}{2} y + \frac{\sqrt{3}}{2}z ,-\frac{\sqrt{3}}{2}y + \frac{1}{2}z))$\\
\hline
$S$ & $(x,y,z)\mapsto (y,z,x)$ cyclic permutation of axes $\Delta=+1$\\
\hline
$S2$ &$(x,y,z)\mapsto (z,x,y)$ cyclic permutation of axes $\Delta=+1$\\
\hline
$C$ & $(x,y,z)\mapsto (-x,-y,-z)$ reflexion from the beginning $\Delta=-1$\\
\hline
$Ex$ & $(x,y,z)\mapsto (-x,y,z)$ cycling, reflexion from planes $yz, zx, xy$ $\Delta=-1$\\
\hline
$Sx$ & $ExX4=C(X4)^{-1}: (x,y,z)\mapsto (-x,z,-y)$ cycling; rotation round axes $x,y,z$ on $90^{\circ}$ with the subsequent reflexion from a plane, a perpendicular axis $\Delta=-1$\\

\hline
\end{tabular}
\end{table}

\begin{table}
\caption{Systems of forming crystallographic groups }
\begin{center}
\begin{tabular}{|l|l|l|l|}
\hline
 \multicolumn{2}{|c}{System}	 &        & System of generatrixs \\ 
\hline
Triclinic  & 1 & $C_2$ & $C$\\ \cline{2-4}
	       & 2 & $C_1$ & It will not be transformed\\\cline{2-4}
\hline
Monoclinic & 3 & $C_{2h}$ & $C, Z_2\equiv C, Ez\equiv Z_2, Ez$\\\cline{2-4}
		   & 4 & $C_{1h}$  & $Ez$\\\cline{2-4}
		   & 5 & $C_2$      & $Z_2$\\\cline{2-4}
\hline
Rhombic    & 6 & $D_{2h}$  & $C, Z_2, X2\equiv C, Ex$\\\cline{2-4}
		   & 7 & $D_2$	   & $Z_2$, $X2$, $(Y2)$\\\cline{2-4}
		   & 8 & $C_{2v}$  & $Z_2$, $Ex$\\\cline{2-4}
\hline
Trigonal 1 & 9 & $D_{3d}$  & $C$, $Z_3$, $X2$\\\cline{2-4}
		   & 10& $D_3$     & $Z_3$, $X2$\\\cline{2-4}
		   & 11& $C_{3v}$  & $Z_3$, $Ex$\\\cline{2-4}
\hline
Trigonal 2 & 12& $C_{3i}$  & $C$, $Z_3$\\\cline{2-4}
		   & 13& $C_3$     & $Z_3$\\\cline{2-4}
\hline
Tetragonal 1&14& $D_{4h}$  & $C$, $Z_4$, $X2 \equiv C$, $Z_4$, $Ex$\\\cline{2-4}
	        &15& $D_4$     & $Z_4$, $X2$\\\cline{2-4}
	        &16& $C_{4v}$  & $Z_4$, $Ex$\\\cline{2-4}
		    &19& $D_{2d}$  & $Sz$, $X2$\\\cline{2-4}
\hline
Tetragonal 2&17& $C_{4h}$  & $C$, $Z_4$\\\cline{2-4}
            &18& $C_4$     & $Z_4$\\\cline{2-4}
			&20& $S_4$     & $Z_4$, $Sz$\\\cline{2-4}
\hline
Hexagonal 1.&21& $D_{6h}$  & $C$, $Z_6$, $X2$\\\cline{2-4}
		    &22& $C_{6h}$  & $Z_6$, $X2$\\\cline{2-4}
		    &23& $D_{6h}$  & $Z_6$, $Ex$\\\cline{2-4}
            &24& $D_{3h}$  & $Z_6$, $X2$, $Ez$\\\cline{2-4}
\hline
Hexagonal 2.&25& $C_{6h}$  & $C$, $Z_6$\\\cline{2-4}
		    &26& $C_6$     & $Z_6$\\\cline{2-4}
			&27& $C_{3h}$  & $Z_6$, $Ex$\\\cline{2-4}
\hline
Cubic 1.    &28& $O_h$     & $C$, $X4$, $Y4$, $(Z_4)$\\\cline{2-4}
		    &29& $O$       & $X4$, $Y4$, $(Z_4)$\\\cline{2-4}
            &30& $T\alpha$ & $Sx$, $Sy$, $(Sz)$\\\cline{2-4}
\hline
Cubic 2.    &31& $T_h$     & $C$, $X2$, $Y2$, $(Z_2)$, $S$\\\cline{2-4}
	        &32& $T$       & $X2$, $Y2$, $(Z_2)$, $S$\\\cline{2-4}

\hline
\end{tabular}
\end{center}
\end{table}
	
With the help of the algebraic transformations (Tables 5 and 6) it is possible to realize the transformations for all 230 crystallographic groups in $S^3$, if one considers a translational component of those groups. Based on the interpretational relation (Table 3) defined in Section\ref{s21}, the following conclusions can be drawn:
\begin{enumerate}
\item[1.]From propositions 1, 2 (Section\ref{s21}) it follows that the transformations $I$,$X2$, $X3$, $X4$, $X6$, $S$, $S2$, $C$, $Ex$ and $Sx$ are completely realized by means of spherical motions of the group $S^3$.
\item[2.]Based on the existing connection between a group of the spherical motions $S^3$ and groups of the elliptic motions $V^3$, as well as the interpretative relation (Table 3), we can assert that the transformations $I$,$X2$, $X3$, $X4$, $X6$, $S$, $S2$, $C$, $Ex$ and $Sx$ are realized in the elliptic space entirely as a set of the orthogonal transformations  $U(g)$ and a set of the paratactic shifts $\delta(g)$.	
\item[3.]Based on the interpretational relation given in Table 3, the crystallographic classes have an elliptic representation on the basis of the orthogonal transformations $U(g)$.
\end{enumerate}

Based on the conclusion 3 and the established interpretational relation (Table 3), any crystallographic group can be represented as a set of the elliptical motions.
Thus, it was found that all 230 crystallographic (Fedorov) groups used to describe crystal structures in a Euclidean space can be represented by elliptical motions of the closed space $V^3$.

\pagebreak
\section{The $I_E^R$-functional. The interpretational \\ relation to represent  of images \\ $F$-groups in $E^3$}

The discrete groups applied in classical physics and acting in a Euclidean space, as groups of transformations of symmetry: crystalline structures, particles, different physical fields and other, even more often need representations and descriptions of the conditions realized in nonlinear spaces, what is dictated by a non-compliance between used realizations of models and physical observations.

The structures of crystallophysical properties and phenomena where the behavior of surface (layerwise) particles is not essential or answers the boundary conditions, can be described as a translational closed finite system, explanations for which serve different technical agreements (von Kármán condition) as well as different representations of crystallographic groups in non-Euclidean closed spaces. One of the alternatives to the traditional approach, where crystals are considered in the three-dimensional Euclidean space, is the model of a crystal, offered by S.V. Rudnev, which is based on the elliptic geometry \cite{bib:b24}.

The internal space of a real crystal is supposed to be satisfying the geometry of the bounded, closed space with the elliptic metrics.The consideration of realizations of crystallographic groups in $V^3$ and the visual interpretation of lattices in $E^3$, developed by the descriptive-geometric method ~\cite{bib:b21, bib:b22, bib:b24, bib:b26, bib:b28, bib:b44}, have been put in a substantiation of the model. We will assume that the external material medium is a differentiated manifold which is homeomorphic to $E^4$. We will denote the medium points by $\xi$, and the space points by $x$. Let $\Phi$ be some fixed smooth embedding of the medium in $R^4$---$\Phi$: $\xi\mapsto x=\Phi(\xi)$. Under the assumption, there is an inverse diffeomorphism $\Phi^{-1}: x\mapsto \xi=\Phi^{-1} (x)$. They say that the value of  $\Phi$ determines the external geometrical state of the medium. For the value of  $\Phi$, a Lagrangian system of co-ordinates $\xi^\lambda $, connected with the crystalline medium and an Eulerian system of co-ordinates $x^i$ are introduced. Then
\begin{equation}\label{eq22}
\Phi^{-1}: x^i\mapsto x^i (\xi^\lambda)
\end{equation}
\begin{equation}\label{eq23}
 \Phi \quad   : \xi^\lambda\mapsto\xi^\lambda (x^i)
\end{equation}
A basic geometrical characteristic of the model of th eexternal medium is the metrics which is determined by the standard distance between points of $E^4$, in which there are corresponding points of the crystalline medium. The set of all the internal characteristics of a crystal determines the internal state or internal geometry of a medium. The internal geometry of a medium is set by the internal elliptic metrics:
\begin{equation}\label{eq24}
ds^2=\frac{4\sum_{i=1}^4 d(\xi^i)^2}{(1+\sum_{i=1}^4 d(\xi^i)^2 /\rho^2)}
\end{equation}
In a general case, the construction of the model of a crystalline medium includes the task of some crystallographic group ($\Phi$), acting in the elliptic space, called a $F$-group of the internal symmetry of a crystal. It is meant that the crystal particles, as well as the connected groups of particles, possess some internal structure that is modeled in the space $V^3$. It also influences upon the motions of particles; so at the motion from a point $x$ of the model to a point y on a curvilinear trajectory (geodetic), the particle comes to a point, generally speaking, with a changed vector of the internal state which has been reached thanks to the transformation from the group $F$. 

It is known that the two-dimensional elliptic space (a plane) cannot be differentially immersed in $E^2$, and cannot be differentially embedded in $E^3$, though supposes the local isometric analytical embedding in $E^3$ (34).
Taking into account the above difficulties which are arising when imaging the elliptic motions and their interpretation, S.V. Rudnev has offered an evident graphic method which allows us to conduct the visualization of different types of the elliptic motions ~\cite{bib:b23, bib:b24, bib:b25, bib:b49}.
Proceeding from the correspondence established in Table 3, as well as considering a general view of the elliptic motions\eqref{eq11}, the crystallographic groups acting in the elliptic space, can be presented by those acting in the Euclidean space:
\begin{equation}\label{eq25}
 T_\pm(\delta_a)g(x)\mapsto e^{\pm 2 \pi bt} \phi(x)
\end{equation}
In the three-dimensional case an action is given by:
\begin{equation}\label{eq26}
(\delta_a,g(x))\mapsto(e^{\pm 2\pi b_1t}\phi(x),e^{\pm 2 \pi b_2t}\phi(x),e^{\pm 2 \pi b_3t}\phi(x))
\end{equation}
where $\phi(x)\in U(g)$, where $U(g)$ is an orthogonal subgroup of rotations $R^4 (S^3)$, $e^{\pm 2 \pi bt}$ is the circular transformations corresponding to paratactic shifts of elliptic space. The crystallographic groups acting in the elliptic three-dimensional space have the following matrix form:

\begin{equation}\label{eq27}
\begin{vmatrix}
\cos \delta_a &sin \delta_a & 0 & 0\\
-sin\delta_a & cos \delta_a & 0 & 0\\
0 & 0 & cos \delta_a & sin \delta_a\\
0 & 0 & -sin \delta_a & cos\delta_a\\
\end{vmatrix}
\times
\begin{vmatrix}
cos 2\pi t & -sin 2\pi t& 0 & 0\\
sin 2\pi t & cos 2\pi t & 0 & 0\\
0 & 0 & 1 & 0\\
0 & 0 & 0 & 1)
\end{vmatrix}
\end{equation}
where  $\delta_a$ is the stationary distance of a paratactic motion on $S^3$.

The modeling space $S^3/\{\pm I\}$ is a closed nil manifold, and the crystallographic groups act as the subgroups $SU(2)\times\tau$, where $\tau$ is an Abelian normal subgroup of the paratactic shifts. In this case we have the following interpretational relation (morphism):
\begin{equation}\label{eq28}
 SU(2)\times\tau\mapsto T^2\times R^1 \quad \mbox{(Open model)}
\end{equation}
\begin{equation}\label{eq29}
 SU(2)\times\tau\mapsto T^2\times S^1 \quad \mbox{(Closed model)}
\end{equation}
Thus, the interpretational pictures are constructed in the Euclidean space that is a priori assumed to be fibered into tori.  The modeling elliptic space, respectively, is fibered into Clifford surfaces $S_K$ which are put in correspondence to the tori of the Euclidean space.  As a base, carrying surface of the interpretation, the Clifford surface $S_K$ has been selected by geometrical and physical considerations. When interpreting, the two-dimensional torus $T^2$ is put into correspondence with each $S_K$, due to the relation established in section \ref{s21} as well as the morphisms (\ref{eq28}, \ref{eq29}). The desired expression for the complete $I_E^R$-functional, which realizes the interpretational relation for of images $F$-groups in the elliptic space $V^3$, is shown in Table 7.

\begin{table}

\caption{The $I_E^R$-functional. The interpretational relation}
\begin{tabular}{|p{4cm}|p{10cm}|}
\hline
Elliptic space ($V^3$) & The space of interpretation ($R_E$)  \\
\hline
The case ~ $n = 1$
$\varphi(x)\in F_1$---a group of actions of an elliptic straight line $V^1$
& 
The space of interpretation is a circle $C_1$, $|z|=1$ on a complex plane (a conformal case), (images) of transformations are considered as a group of complex numbers with multiplication as a group operation. An image of the group $F_1$ is a group of transformations:  $\Psi_1 (x):f(x)=e^{2\pi i\varphi(x)}$, consisting from
$T_0^1=e^{2\pi ia}$ (translations) $T_1^1=e^{-2\pi ia}$ (reflexions).
\\
\hline
The case $n = 2$ $\varphi_2 (x)\in F_2$---a group of actions on a elliptic plane $V^2$
&
A complex plane $C_2$  with 2-dimensional representations $\Psi_2(x)$ acting on it serves as the space of interpretation: 
\begin{tabular}{c c}
$T_0^2=
\begin{vmatrix}
e^{ik\pi a} & 0 \\
0& e^{-ik\pi a}\\
\end{vmatrix}$
&
$T_1^2=  \begin{vmatrix}
0 & e^{-ik\pi a} \\
e^{ik\pi a} & 0\\
\end{vmatrix}$\\
(rotations and translations) & (reflexions)

\end{tabular}
\\
\hline
The case $n = 3$
$\varphi_3 (x)\in F_3$---a group of actions in the elliptic space $V^3$  
&
The space of interpretation is considered in two forms:
\begin{enumerate}
\item[1)]Open space: $T^2\times R^1$.
\item[2)]Closed space: $T^2\times S^1$.
\end{enumerate}
	
The following 3D representations $\Psi_3(x)$ are an image of the groups $F_3$:

\begin{tabular}{c c}
$T_1^3=
\begin{vmatrix}
e^{ik\pi a} & 0  & 0 \\
0 & e^{-ik\pi a}   & 0\\
0  & 0 & e^{ik\pi a} \\
\end{vmatrix}$
&
$T_2^3= 
\begin{vmatrix}
e^{ik\pi a} & 0  & 0 \\
0& e^{-ik\pi a}   & 0\\
0  & 0 & 1 \\
\end{vmatrix}
$\\
(rotations) & (shifts)\\
$T_3^3=
\begin{vmatrix}
e^{ik\pi a} & 0  & 0 \\
0& e^{-ik\pi a}   & 0\\
0  & 0 & -1 \\
\end{vmatrix}$
&
$
T_4^3= 
\begin{vmatrix}
0 & 0  & e^{ik\pi a} \\
0& e^{-ik\pi a}  & 0\\
e^{ik\pi a}  & 0 & 1 \\
\end{vmatrix}
$\\
(shift and reflexion)& (reflexions)\\

$T_5^3=
\begin{vmatrix}
0 & 0  & e^{ik\pi a} \\
0& e^{-ik\pi a}  & 0\\
1  & 0 & -1 \\
\end{vmatrix}$
 & 
$
T_6^3= 
\begin{vmatrix}
0 & 0  & e^{ik\pi a} \\
0& e^{-ik\pi a}  & 0\\
-1 & 0 & 1 \\
\end{vmatrix}
$\\
(reflexions and shift) & (reflexion and rotation)
\end{tabular}

\\
\hline

\end{tabular}
\end{table}

\pagebreak
\part*{Results and discussion}

At the present time, it has become increasingly clear that the interpretation will result not only in the visualization of some geometric images with the help of others, but also that, according to the definition of interpretation, it is the way to implement the features of one geometric system within the other. This is especially important when considering models of physical processes and phenomena in the space of interpretation.

We can assume that none of the basic geometric systems ($S^3,H^3,E^3$) can be realized at all as the basis for the real crystallographic space---that is, as a single geometric system for the space of our experience.

We assume that the difficulties arise mainly when applying a common geometric system (one of $S^3,H^3,E^3$) to different real-world objects. In the real world, at different levels of organization of matter, there is a realization of some geometric systems and forms of their interaction within other geometric systems in accordance with the geometric features of each---that is the realization of: elliptic geometry in Euclidean space, Euclidean geometry in Lobachevsky space, and so on. In this system, there is no hierarchical subordination, but only a mutual realization of one geometric system within the other, in accordance with the laws of interpretation. The proposed approach does not conflict with quantum-mechanical ideas about the structure of atomic and ionic particles forming the crystal.

The chosen approach corresponds to R.V. Galiulina's thesis \cite{bib:b20} about the relationship of axiomatics of the geometric apparatus and the geometrical features of the crystallographic space. As a modeling space which describes the internal structure of a crystal, the closed space $V^3$ with the elliptic metric and constant positive Gaussian curvature $(K=1)$ is proposed. To visualize the results, we use the specially developed interpretation of geometric objects  ($S_K$,  $F$-groups, symmetries) of the modeling space $V^3$ using the constructions in a Euclidean plane $E^2$ and space  $E^3$. All the aforesaid allow us to assert with all certainty the following:

\begin{enumerate}
\item[1.] Based on the analysis of discrete groups of motions of the spaces $V^3$ of positive curvature $(K =1)$, the interpretational relation for the modeling space has been established. Because Clifford translations of the elliptic space are isomorphic to the ordinary translations (parallel translations) of the Euclidean space, it allows us to build and justify the interpretation of Euclidean crystallographic groups using the discrete groups of elliptic motions.
\item[2.] Using the algebraic and matrix representations of crystallographic \linebreak groups, in the present work it has been shown that all 230 crystallographic groups used to describe crystalline structures in the Euclidean space can be represented by elliptic motions in the closed space $V^3$. For any crystallographic group one can construct an elliptical representation using the group of elliptic motions.
\item[3.] A model of a closed, compact crystal structure in $V^3$ has been proposed, and the ways of interpretation of its elements of symmetry of the crystal lattice in the Euclidean space $E^3$ have been identified.
\end{enumerate}

An important difference in the proposed model constructions is a more complete (accurate) account of the structural features of crystal structures at different scale levels, including the sources of the formation and evolution of curved regions (defects). Based on the results, it can be argued that the crystallographic space of interpretation  $R_E$---a special geometric system---can serve as the geometric model of the real crystallographic space.

\pagebreak

\end{document}